# Conceptual Design Of An Ideal Variable Coupler For Superconducting Radiofrequency 1.3GHz Cavities.

Chen Xu, Sami Tantawi

Stanford Linear Accelerator Center, Stanford University, 2575 Sand Hill Road, Menlo Park, California 94025, USA
Electronic address: chenxu@slac.stanford.edu

## Abstract

Inspired by the development of over-moded RF component as an undulator, we explored another over-moded structure that could serve the variable coupling for SRF purpose. This application is to fulfill variation of S11 from 0 to -20db with CW power of 7 KW. The static heat loss in the coupler is trivial from calculation. An advantage of this coupler is that the thermal isolation between the 2K and 300K section is considerable by vacuum separation. Within this coupler, only a single propagation mode is allowed at each section, and thus, the fact that no energy is converted to high order mode bring almost full match without loss. The analytical and numerical calculation for a two window variable coupler is designed and optimized. A RF power variation is illustrated in the scattering matrix and coupling to cavity is also discussed.

## 1. Introduction

Superconducting radiofrequency technology is widely implemented in various accelerators and collaborations around the world. Superconducting cavities and input couplers have been optimized in Tesla collaboration for two decades. TTC-3 coaxial type input coupler is optimized for different purposes through three generations. Two of major optimizations goal are reducing the static heat loss and minimizing electric field on the vacuum windows. Here we design a new over-moded input coupler structure to reducing the static heat loss.

The TTC -3 type coupler is a variable coupler. At 2K, it generates static heat loss 0.02 W and the dynamic heat loss is 0.06W at transmission average power level of 2KW. It has three cambers and each camber has its own vacuum. But basically, it consists of three parts: a door nob: waveguide to coaxial transition, center vacuum impedance match and the top tip antenna to the cavities.

1. The main static loss of this coupler comes from the door-nob transition. It is good for a SRF pulse application, however for a CW operation, one needs to reduce the static losses.

2. Second, whole couplers are cooled with conduction loss without any external cooling. The tip top part has the highest temperature in this coupler, the peak can go up to several hundred K. This heat needs to transfer out to the base part where nitrogen can take the heat away. Thus, there would be a temperature gradient from the tip to the nitrogen cooled outer conductor and this steep gradient will cause metal surface thermal stress and compromise the mechanical tolerance.

3. Introducing the bellow can help reduce the thermal stress and introduce the movable mechanism. The bellows attached both inner and outer conductors bring the manufactory difficulties and mechanical alignment through all parts.

4. Waveguide couplers do not have inner conductors, and they are easy for manufacturing and cooling. However, they usually offer no variable mechanism.

Coaxial couplers such as like TTC3 coupler conduct RF power in a coaxial TEM mode. But first one needs to convert the rectangular waveguide TE10 mode from a klystron into coaxial TEM mode by a door-nob structure. Then, the coaxial line will transmit the RF power with various ohmic impedance matchers to an appropriate radius combination on the cavity beam pipe receiver. On the other hand, waveguide couplers usually conduct RF power in form of the TE10 mode, which is the fundamental mode of a rectangular waveguide. Once a waveguide is manufactured, it is hard to change the coupling online. Reflection from both sides should be less than -30db through the variable length.

In this paper, we introduce a new input coupler without physical contact from the input waveguide and cavity receiver. The structure can convey RF power at level of 7KW without calculated multipacting barrier, while produces equivalent dynamic loss and negligible static heat loss. The whole coupler shown in Fig. 1 can be divided into three sections. Left part is the transition from RF source to sapphire loaded converter. Middle is the sapphire guided transmission line. Right part is a convertor to cavity with a given iris.

Fig.1. HE11 mode coupler in a final mechanical drawing.

## 2. General consideration:

**2.1 Introduction of Hybird modes:**

Hybrid modes are first labelled by Snitzer in 1961, and it is more universal mode than the standard TE/TM modes. These modes have both Hz and Ez components and TE and TM modes come together.

In a modern optics fiber industry, HE11 mode is widely used for convey signals. A simple dielectric rod can be used for RF transmission. Inside of this rod, fields can descripted by a series of Bessel functions, while fields outside of the rod are usually evanescent modes in radial direction and surface modes in Z direction. The evanescent mode is the radiation mode with continuous spectrum, and the outside fields can be depicted by modified Bessel functions. In the Z direction, both inside and outside fields should continue on the dielectric and vacuum interface. The structure is shown in fig.2.

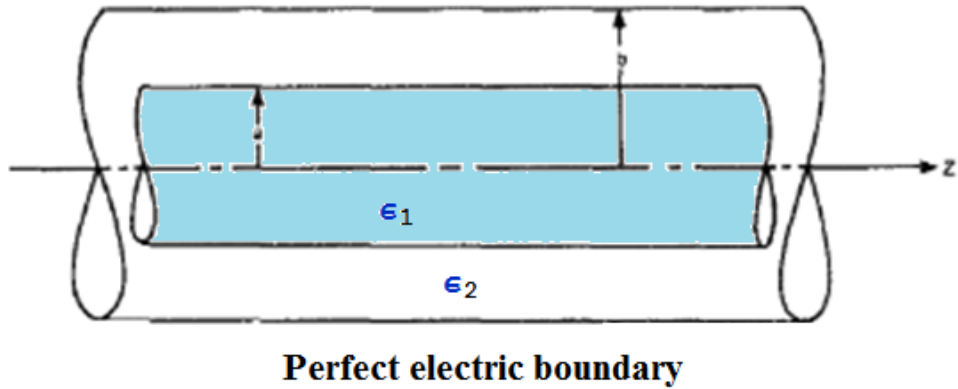

Perfect electric boundary

Fig.1. The sapphire loaded waveguide.

Presuming the rod radius is $a$, without $e^{-j\beta z - im\phi}$ term, the field in sapphire rod and between outer conductor is described:

*When* $\rho \geq a$

$$E_z(m, \beta_s]) = A\, K_m(\alpha_s \rho)$$
$$\eta_2 H_z(m, \beta_s]) = B\, K_m(\alpha_s \rho)$$

*When* $\rho \leq a$

$$E_{zs}(m, \beta_s]) = C\, J_m(\chi \rho)$$
$$\eta_2 H_{zs}(m, \beta_s]) = D\, J_m(\chi \rho)$$

Where $K_m$ is modified Bessel function of the second kind, J is Bessel function of the first kind, $\eta_2$ is the wave impedance, $\alpha_s, \chi$ are the wave numbers in the sapphire and stuffed material and A,B,C,D are the amplitudes of the field. Other transverse fields can be calculated from:

$$E_t = \frac{-i\omega\mu}{\lambda^2}[\nabla_t H_z \times \hat{z} + (\frac{\beta}{\omega\mu})\nabla_t E_z]$$

$$H_t = \frac{-i\omega\mu}{\lambda^2}[\hat{z} \times \nabla_t E_z + (\frac{\beta}{\omega\mu})\nabla_t H_z]$$

Where $\lambda$ is the wavelength of propogating wave.

Moreover, the relation of the wavenumber on each side is:

$$\beta_s^2 + \chi^2 = k_0^2 n_1^2$$
$$\beta_s^2 - \alpha^2 = k_0^2 n_2^2$$

Where $n_1$ $n_2$ are the reflection indexes in each material and $k_0$ is the free space wavenumber.

On the sapphire interface, $E_z$ $H_z$ $E_\phi$ $H_\phi$ are continuous. We can derive the equations from both the sapphire and vacuum sides, and the 4× 4 equations have a determinant must be zero in order to obtain a nontrivial solutions. For those fields, the modal equation for wavenumber $\beta_s$ is written as:

$$[\frac{J'_m(u)}{uJ_m(u)} - \frac{K'_m(v)}{vK_m(v)}][n_1^2 \frac{J'_m(u)}{uJ_m(u)} - n_2^2 \frac{K'_m(v)}{vK_m(v)}] = (\frac{m\beta}{k_0})^2(\frac{1}{v^2} + \frac{1}{u^2})^2$$

$$u = \chi a$$
$$v = \alpha_s a$$

In these functions, if m is zero, the left side breaks into $\frac{J'_m(u)}{uJ_m(u)} - \frac{K'_m(v)}{vK_m(v)}$ for a simple TE mode or $n_1^2 \frac{J'_m(u)}{uJ_m(u)} - n_2^2 \frac{K'_m(v)}{vK_m(v)}$ for a TM mode. If m is not equal to zero, a series of hybrid modes are developed. These Z-direction propagation intertwined modes are surface wave modes that have a discrete spectrum and each mode has a cutoff frequency whose wavenumber approaches zero. For lowest mode, the HE11 mode has a cutout frequency for zeroth root of $J_1(V)$, where the normalized frequency is $V = k_0 d(n_1^2 - n_2^2)^{1/2}$. For this reason, there is no cutoff for the HE11 mode. Higher nth order hybrid modes have cutoff frequencies at nth root for $J_1(V)$. The HE11 mode has field pattern shown in Fig.2.

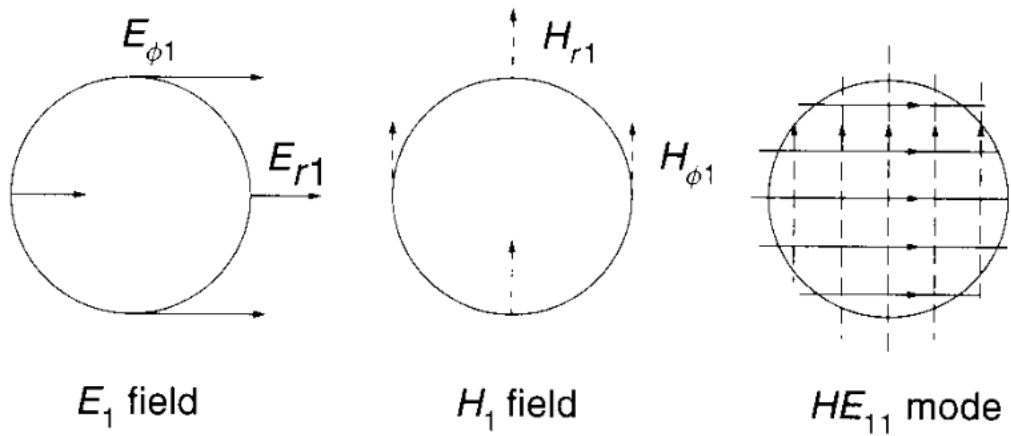

Fig.2. Field pattern of HE11 mode, courtesy from Elements of Photonics, Volume II.

In optics research, since the field inside of core materials is pure linear in the transverse plane, it is also called linear polarizer (LP) mode. LP01 mode is the same as HE11 mode. The next higher mode, LP11, is shown in Fig.3 with clad material around the core materials.

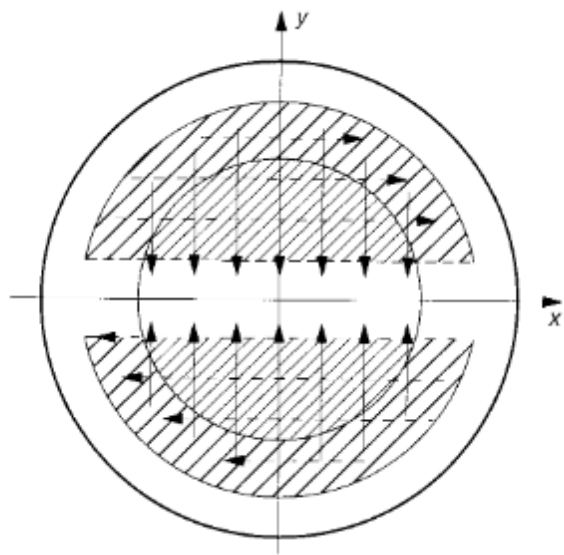

Fig.3. Field pattern of LP11 mode, courtesy from Elements of Photonics, Volume II.

Note, the fields are oscillatory in the core and evanescent in the clad materials. The decay rate outside the rod has an upper bound $\leq k_0\sqrt{(n_1^2 - n_2^2)}$, when $\beta_s$ is zero. To get the larger decay, larger $n_1$ is required. For this reason, sapphire $\varepsilon$ (~10) is chosen where n is around 3.3 and loss tangent is fairly low. This also means that if one can cleverly choose $\beta_s$, the power decays fast outside of rod, thus the fields power are highly guided by the rod. In the optical case, a pair of reflective index is chosen to close to each other in order to get a weakly guided power. Note the wave phase velocity inside the rod is less than that of place waves in the free space.

**2.2 Fundamental Design options:**

As mentioned above, $\beta_s$ should be chosen small in order to get the quickest decay outside of the rod. By tapering the radius, one can change the $\beta_s$. Reducing the radius can increase the cutoff frequency of the propagating modes. To calculate the cutoff frequency of HE21 mode which is the next higher mode, the radius is expressed:

$$d = \sqrt{\frac{V^2}{(n_1^2 - n_2^2)k_0^2}} = \sqrt{\frac{c^2 V^2}{(n_1^2 - n_2^2)\omega^2}}$$

where $V$ is the second root for $J_1(V)$ and is 3.84 and $k_0$ is wavenumber for 1.3Ghz. Note HE11 has no cutoff frequency.

Radius d in this equation should be small enough to prevent HE21 mode has larger cutoff frequency than 1.3 GHz. Within the radius limit, a small diameter rod will result in that the field extends for considerable distance beyond the surface transversely. A radius is optimistic for obtaining highly guided HE11 mode, if beyond, HE21 is propagating, if below, the field is weakly guided. One can freely change the radius, in order to change the power distribution inside and outside of the rod.

To connect to the klystron power source, one needs to first convert TE10 WR 650 rectangular waveguide with a simple TE11 circular waveguide. The radius of this circular waveguide is chosen to accommodate only one TE11 mode. This circular waveguide would be the outer conducting wall of this sapphire load structure. Field power is transmitted into rod HE11 mode within the first taper rod. A dielectric rod with carefully chosen radius ratio of waveguide and dielectric rod confines the field inside its body rather than a waveguide. Once the power is highly guided in the rod so that the outside waveguide can by tapered down and seal on the sapphire for vacuum purpose. On the receiving side, both the rod and outside waveguide are tapered in order to push the power in rod into the TE11 mode into the waveguide. Since the RF waveguides from both ends don't contact each other, there is little static heat transfer, nor is the high temperature gradient. Meanwhile, a variable matching mechanism can be easily adapted when these two ends are movable respectively towards each other. Once the power is fully converted into the TE11 mode in the receiving end, one can simply convert this circular mode into coaxial mode to feed the cavity without changing the pre-existing main input coupler port on the beam axis. Note, for the vacuum isolation concerns, one can add another ceramic window inside coaxial cable.

### 3. Simulation

**3.1 Simulation by parts:**

A superconducting radio frequent resonator is usually operating in L band, for example in our case 1.3GHz. A WR 650 rectangular waveguide conveys the power to the klystron and converts into the circular waveguide, which supports the TE11 mode with one polarization. This converter has been developed and widely used elsewhere. We use a circular waveguide of radius 100mm as outer conductor and a sapphire rod with redius of 25mm as inside conductor. The electromagnetic field pattern of the HE11 mode are calculated by HFSS software and illustrated in Fig.4. There is neither TE 11 mode nor TEM mode in the waveguide at 1.3 GHz. The scattering parameters are also shown in Fig.4., when the coaxial is matched. Other higher order modes are TE and TM waveguide guide modes and not sapphire guided modes.

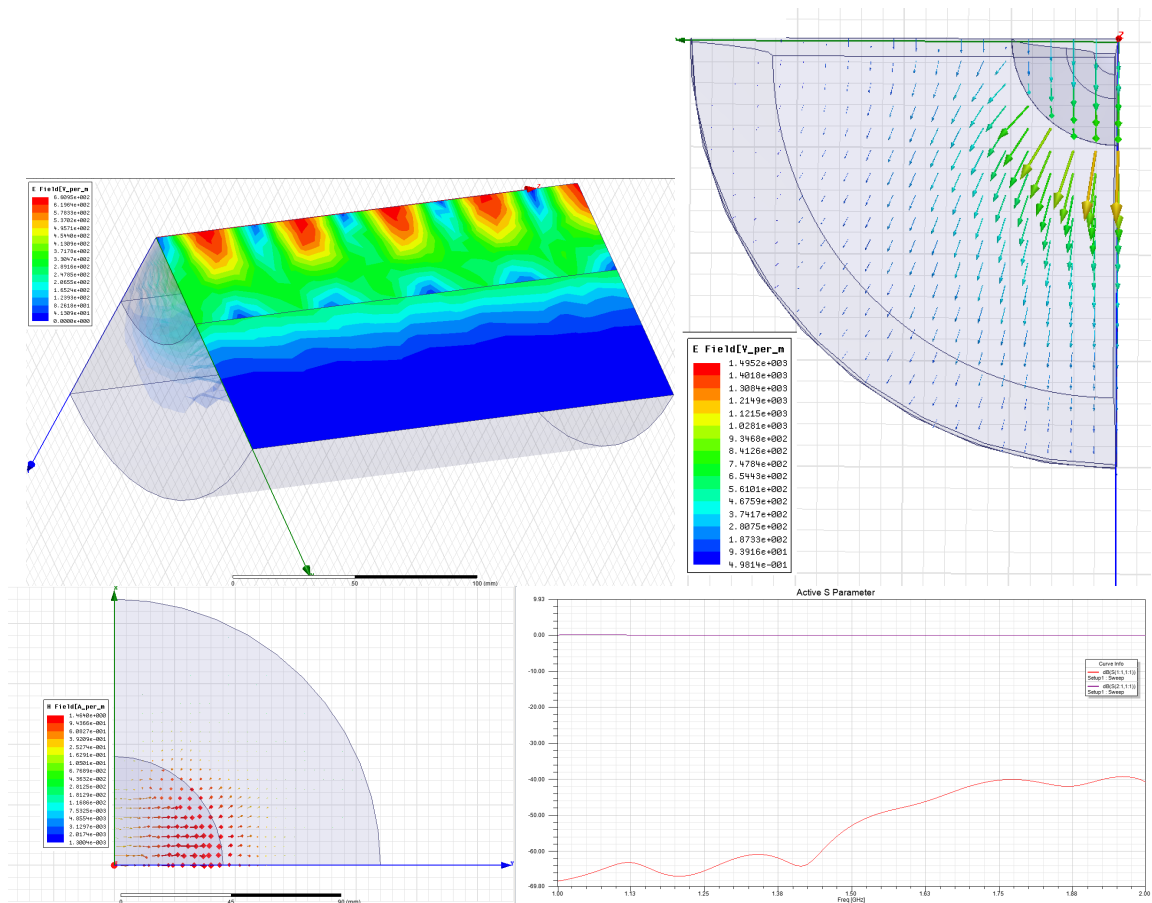

Fig.4. Field pattern of HE11 mode in the center transmission section.

The field decay outside of the rod can be altered by varying the radius by equivalently changing the $\beta_s$. A smooth transition is required to convert the circular waveguide to the HE11 sapphire loaded structure. Higher order modes are introduced to match the field patterns at the discontinuous interface. However, those modes are evanescent modes, which cannot propagate outside with a given length. A double arc structure is implemented here to achieve smooth transition. The configuration is shown in Fig.5.

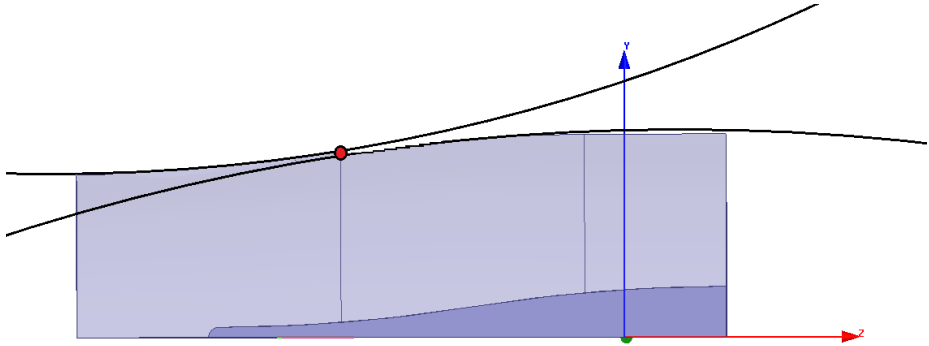

Fig.5. Illustration on double arc structure: the position of red dot is one optimization parameter within given length. Two solid lines are two parts of two circles and indicate the baseline of outer conductor, and at the red dot, they are circumscribed.

This double arc structure design is implemented on both the inner and outer conductors, and the circumscribed positions are the optimizing parameters for obtaining a match. Within a limited given transition length and angel, the position of the red dot is optimized to obtain a match. By doing this optimization, the power of TE11 in waveguide is pressed into sapphire rod body.

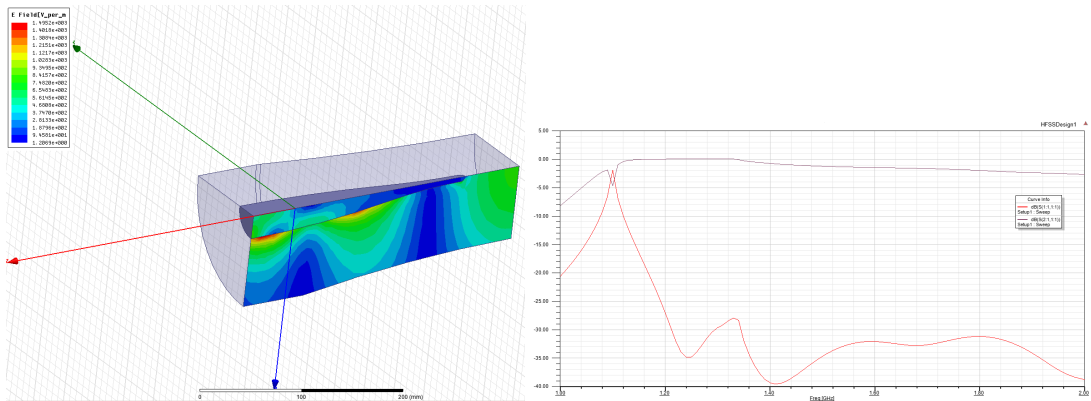

Fig.6. Field pattern and scattering parameters in the converter section.

After the TE11 mode is converted into the HE11 mode in sapphire, the field decays transversely beyond the surface. The outer conductor has almost no effect on the strong guided power, so we manage to taper the outer conductor down onto the surface, and thus, the sapphire rod acts like a window for vacuum purpose. In Fig.7, the sapphire rod (dark blue) is tapered up and the outer conducting waveguide (light blue) is tapered down by two double arc structures. On the positive z-axis, there is no outer conductor anymore and the power is guided by the sapphire rod. However, for mechanical support and alignment reasons, two copper plates are added inside of XY plane and cavity side waveguide. This is the open to the vacuum area and also a section breaking the RF source and cavity receiver. An outer conductor is added to attach to the cavity wall and these two plates forms a gap open to cryogenic vacuum. These two plates must be long enough to ensure no RF field leakage even there is electromagnetic at this gap in the final design. The structure configuration is demonstrated in Fig.7. There are three optimization parameters: two transition points for sapphire and outer conductor and final sapphire rod radius. This discontinuity in outer conductor will bring an advantage for the static thermal study.

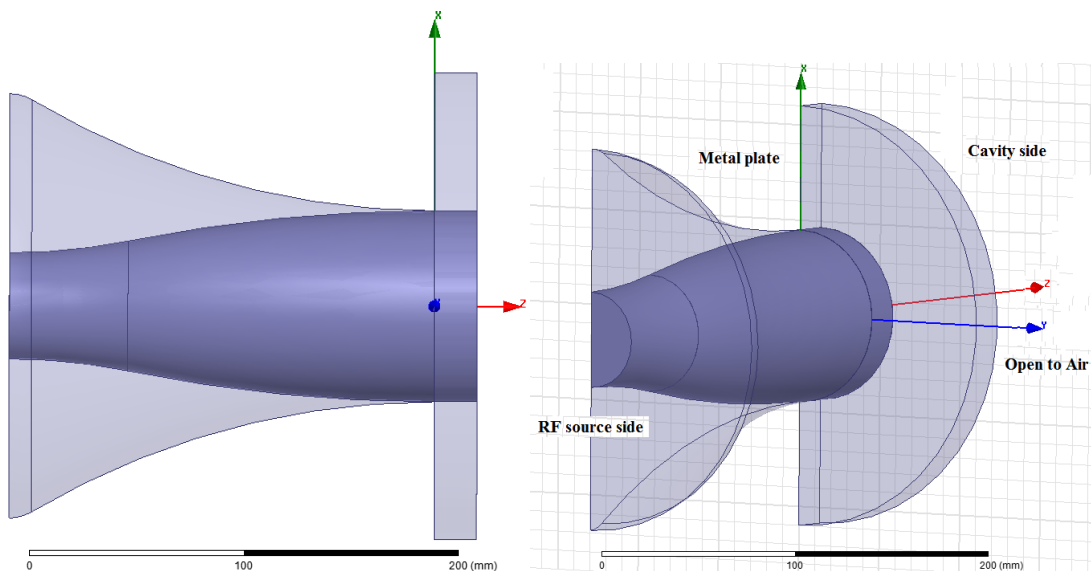

Fig.7. Field pattern and scattering parameters in the converter section.

Scattering parameters are scanned for different frequencies and the match is around -30db at 1.3 GHz in Fig.8. In finite element method (FEM), one has to determine a boundary condition, whereas air in Fig 7 is not a supported boundary condition unless assigned as a wave port. However, we predict there is no field at this gap. In this study, we use a perfect match layer, in HFSS, which means that the power incidence to this surface will be absorbed regardless of mode or frequency. It truly represents the real structure to avoid the 77K and 4K RF component attachment.

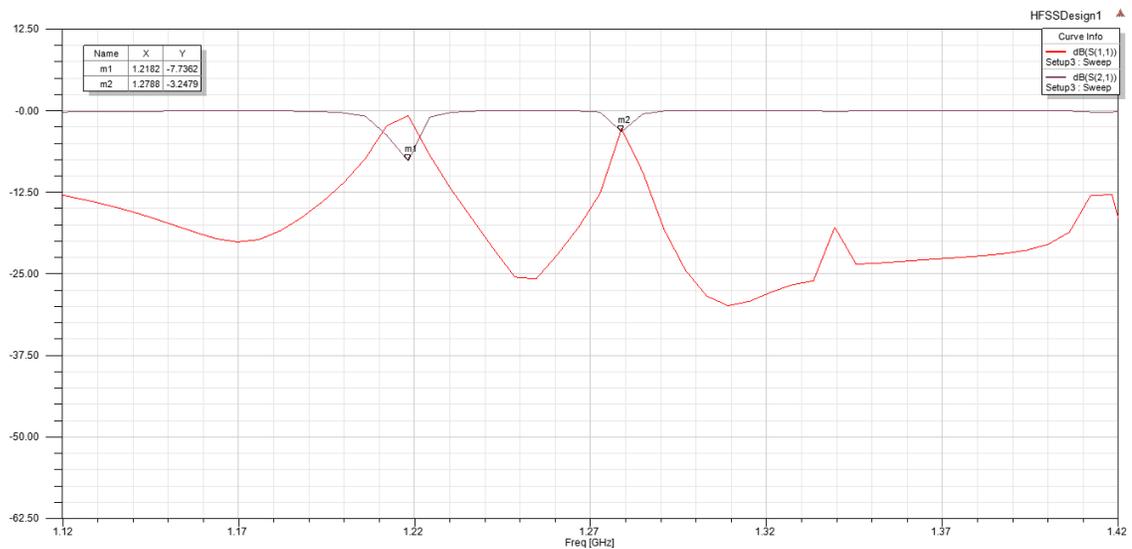

Fig.8. Scattering parameters in the converter section including air perfect match layer.

On the cavity side, similarly, one can continue using this double arc technique but reverse the directions. However, note that one now taper both inner and outer conductor down with one stage or several stages. The advantage is that there will be a fewer number of higher order modes excited and better matching can be obtained. Fig. 9 shows a double arc structure with two stages at the expense of the structures compactness.

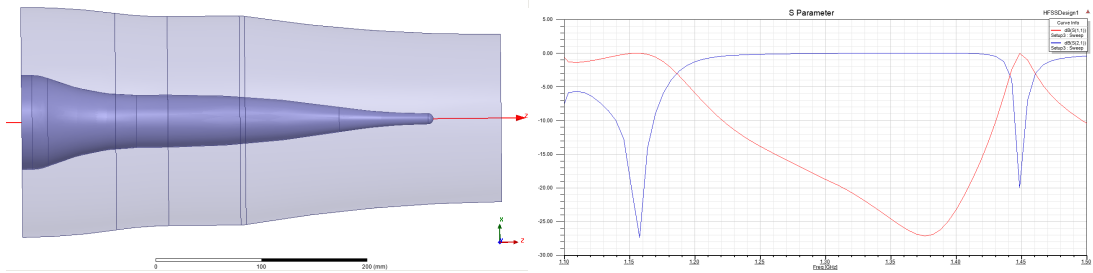

Fig.9. geometry structure and scattering parameters in the cavity converter.

Since the Tesla Test Research (TTR) shape 9 cell cavities are well developed, it is fairly expensive to remodel the end cell and the attached pipe line with a fundamental coupler hole. We would like to design a converter to change circular TE11 mode into the predetermined coaxial aperture on the TTR cavity. The coupler aperture is 40mm and inner conductor diameter is 20mm. On this coaxial cable, there is an ideal position to isolate vacuum from cavity and the coupler. A ceramic windows is added inside the coaxial cable, and the electric field on ceramic windows is minimized by properly increase the inner conductor radius. The windows thickness is chosen to 7mm. It is not necessary to add a choke joint, since the complete structure is in the 4K environment. The structure of this TE11 to TEM are shown in Fig 10.

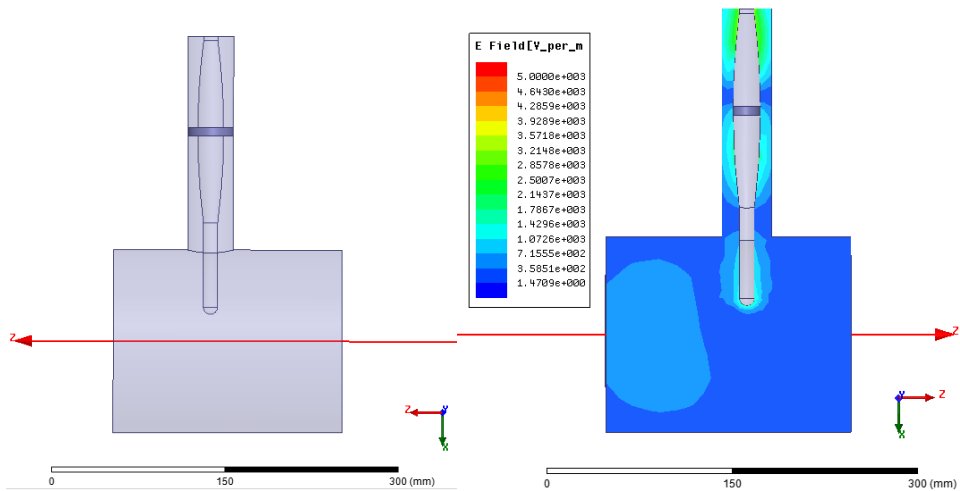

Fig.10. coaxial converter geometry and scattering parameters in the cavity converter.

### 3.2. Final design:

After cascading the parts as described above, one can obtain a full structure of this HE11 mode coupler. This section combination exercise ensures that each subsystem has a good match, however, it might be not the most compact structure. To get a compact design, a full structure optimization is needed. A whole structure geometer is plotted in Fig.11.

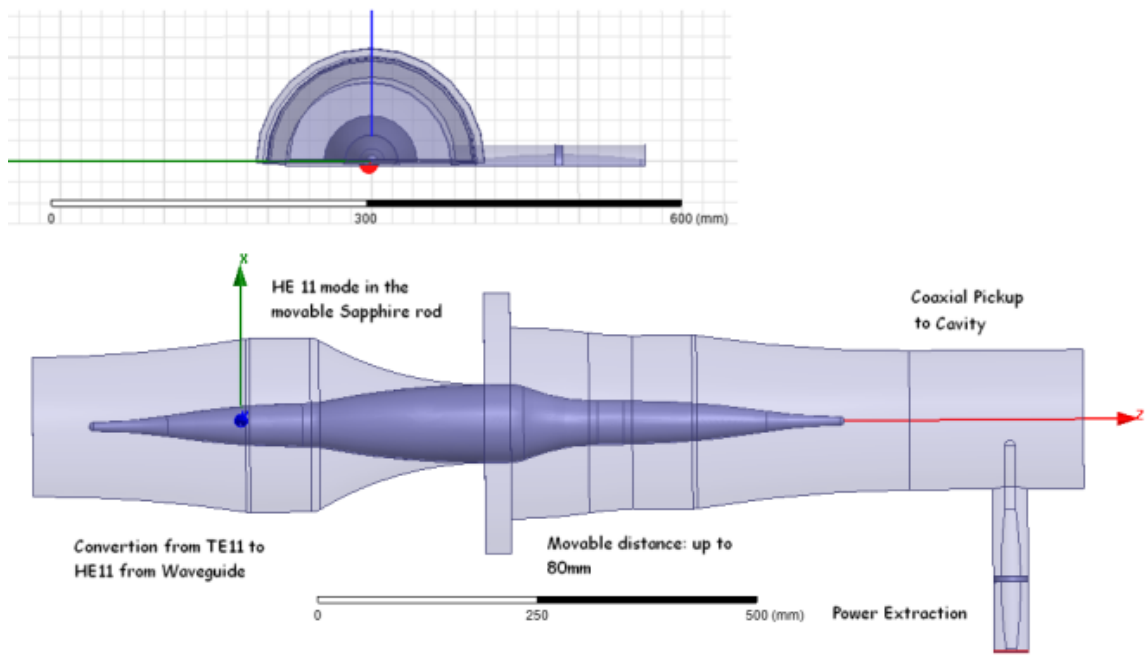

Fig.11. Geometry structure of this HE11 mode power coupler.

As stated above, the center part has an 'open to air' component in which RF power is transferred only in the sapphire rod. Electromagnetic fields are plotted in the Fig.12.

Field distributions in Fig.12 suggest that the field is strongly guided in the sapphire (center). Since strong guided, there is no electric or magnetic field in the 'open to air' components. Thus there is no significant RF power leakage. However, this 'open to air' structure effectively breaks the possible temperature gradient in the outer conductor. RF passed through sapphire rod is pushed out, converted into the TE11 circular waveguide and finally transmitted to the coaxial antenna and thus to accelerating cavities.

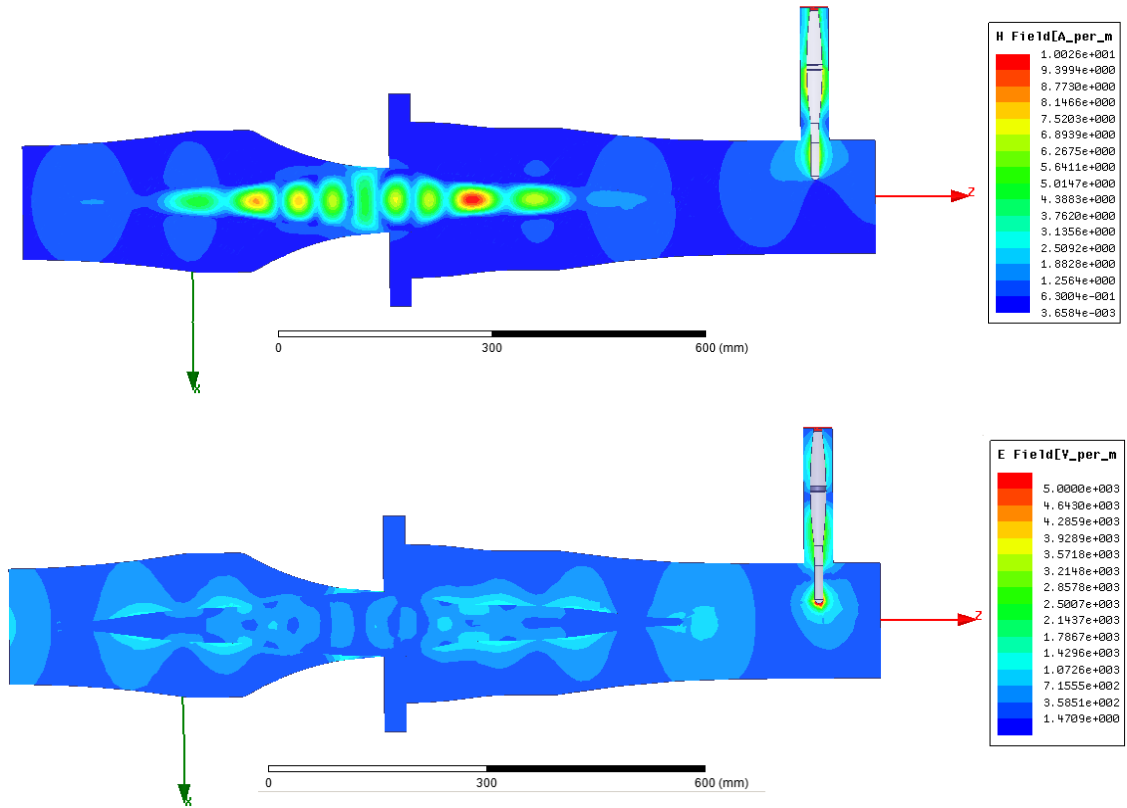

Fig.12. Electromagnetic fields in this HE11 mode power coupler with input power 1W.

The optimized narrowband coupler should enable S11 and S21 a changing range as large as possible at frequency of 1.3GHz. After sweeping the multi parameters above, one needs to obtain a S11 from 1 to 0 in order to achieve full match or full reflection. Fig.13 shows the optimized S reflection and matching parameters at a full matching condition.

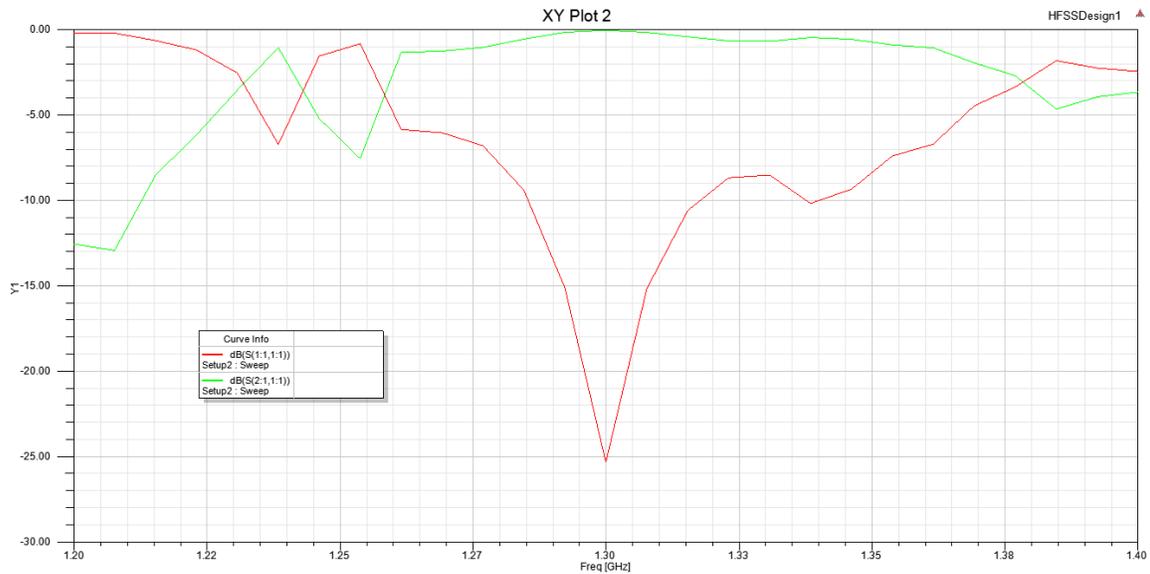

Fig.13. S11 and S21 scattering parameters are shown as a function of frequency with full match condition.

## 4. Other concerns
### 4.1 Variability of match

We have two separate parts in this coupler: one part is connected to the RF source and the other part is attached to the cavity pipeline. The sapphire is held by the tapered outer conductor and acts like a vacuum seal. Between these two parts, there is an opening to the cyro-module vacuum, though one can connect them with a bellow for longer thermal path. This structure leaves us with movable mechanics to change the forward power to the cavity side. This is another method to change the power transmission without changing the coupling of coaxial and cavity as in TTC 3 coupler. The movable geometry is shown in Fig. 14.

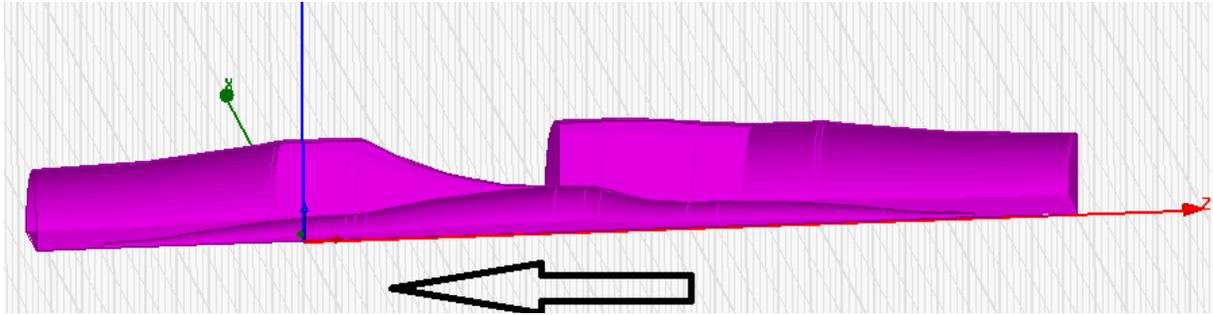

Fig.14. Power coupler moving mechanism.

When the coupler is inserting or extracting, the 'open to air' gap is effectively decreased/increased in length, however, it still produces no RF leakage, because the power is strongly guided by the sapphire rod. The scattering parameters are changed during this movement. The relation of the S parameters and open gap length is shown in Fig. 15.

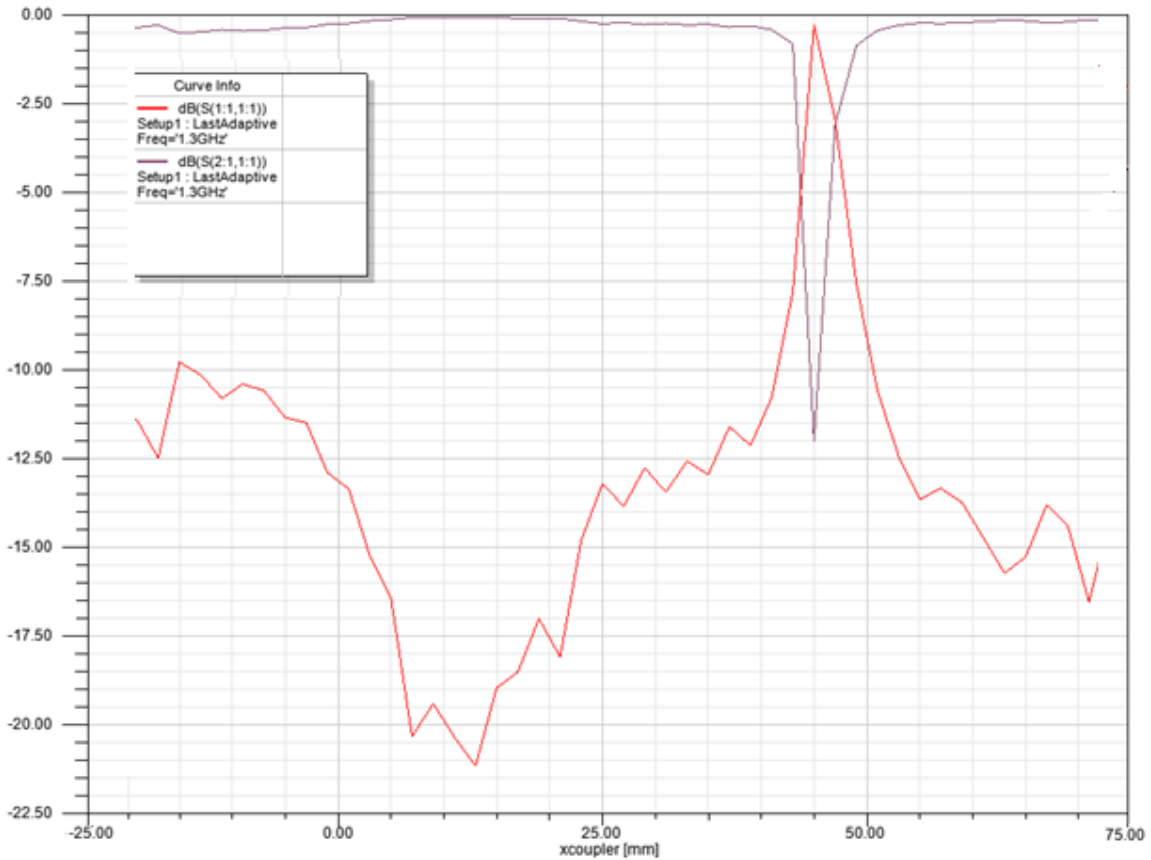

Fig. 15. The relation of S parameters and open gap length.

When the gap is around 12.5mm, the coupler reaches the full match. While the gap is 45mm, the coupler reaches full reflection. Fig.16 shows the frequency spectrum at full reflection.

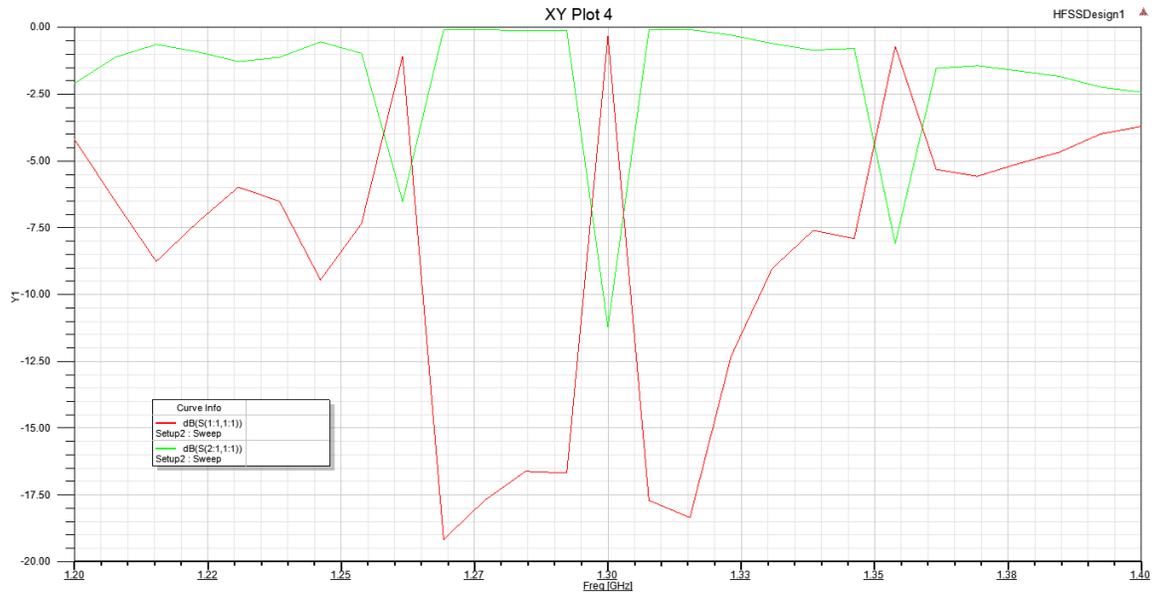

Fig.16. S11 and S21 scattering parameters are shown as a function of frequency with full reflection condition.

### 4.2 Multipacting

Spark 3D can determine the microwave breakdown power level in a wide variety of passive devices. Similar to ACE 3P, it calculates secondary electrons number by following each electrons trajectory. Within a given duration, if the surviving electrons number increases, software consider that there is a multipacting barrier at this field gradient or breakdown. The multipacting study of different power level is given in Fig 17. This result indicates this structure can work with CW power of 7.25kW.

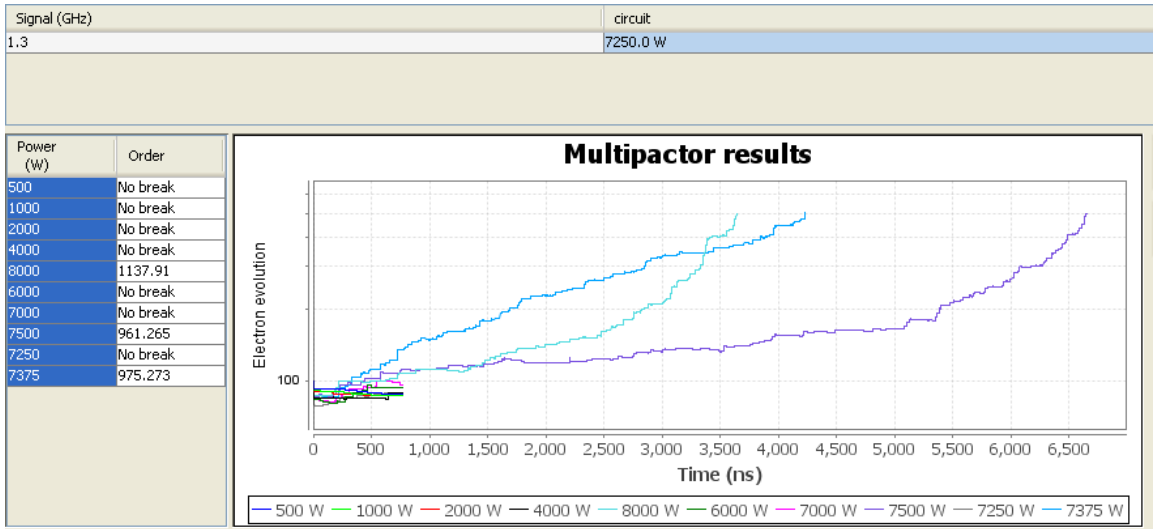

Fig.17. Multipacting study by Spark 3D, this study can show the maximum RF power level this coupler can operate with multipacting.

**4.3 Integrated Heat study:**

The maximum power of this coupler is determined by the maximum surface E field in the design. From Fig. 12, the maximum electric field of $\sim 9.6\times10^3 V/m$ occurred at the tip of the coaxial pickup. This means that if this coupler operate at 7kW, the max E field will be around $803\times10^3 V/m$, which is below the breakdown power for sapphire at 1.3GHz. Beyond the breakdown field, the structure is discharged and power is taken by the discharge. Similarly, the maximum E field happens at the tip top of coaxial antenna inside the cavity side. One can adapt tip top design from the Tesla coupler and use this flat top to achieve ideal impedance match.

We studied the integrated RF heat loss from the outer conductor, presumably copper. The total RF dynamic loss is 6.667 W and the loss on sapphire is $1.26\times10^{-8}W$ at continuous RF power of 7KW at room temperature. The RF heat distribution is shown in Fig 18.

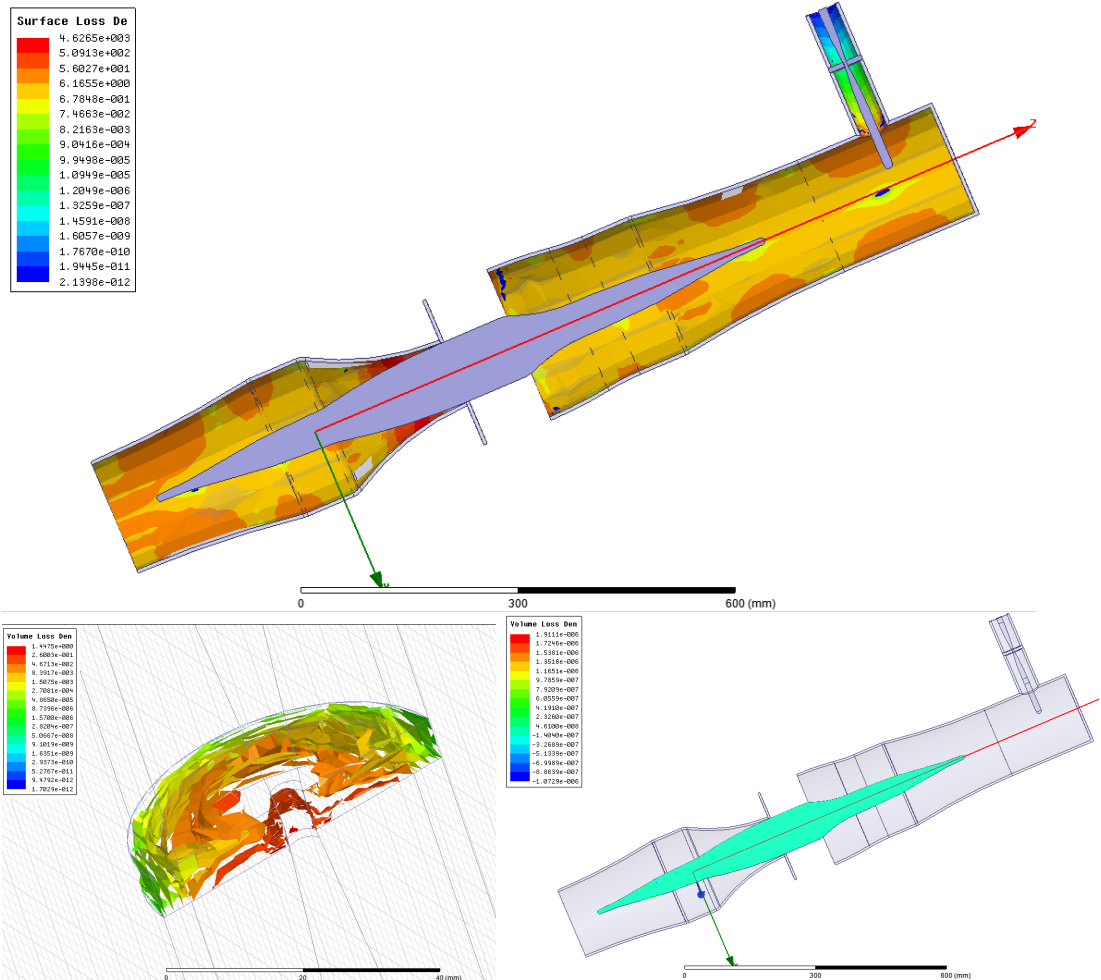

Fig.18. RF power loss on outer conductor is integrated, and total is 6.67W. RF loss in coaxial windows is $7.178\times10^{-8}W$, while sapphire loss is $1.26\times10^{-8}W$ at 7KW level.

**4.4 Ghost modes:**

Since a sapphire rod is located in a circular waveguide and a ceramic window is adapted in the coaxial cable, a trapped mode study requires an eigenmode investigation. Usually, a ghost mode is trapped in a ceramic disk, so dielectric losses of the ceramic increase the temperature. This is one of the major reasons for the puncture and the crack. One will see a resonance of the ghost mode with transmission (X-Y view) measurement. The ghost mode resonance frequency can be moved by adjusting the diameter of the outer conductor and thicknesses of windows.

In our case, a simple RF window used nth quarter wave impedance machining method; a thickness of ceramic disk is in the range of 3- and 3.5-mm. So, a quarter wave length at the RF window part is provided by a pill-box + ceramic disk, not only by ceramic disk.

An Eigen-mode study in table 1 shows that, around working frequency of 1.3 GHz, the nearest neighbor frequencies for this design are 1.27GHz and 1.32 GHz. Their quality factor Q are 22109.6 and 22210.2 respectively. With this narrow band structure with Q external around ~$10^5$, the required bandwidth is 13KHZ, the ghost modes are far beyond consideration.

| Eigenmode | Frequency (GHz) | Q |
|---|---|---|
| Mode 1 | 1.12103 +j 1.52482e-0... | 3.67593e+007 |
| Mode 2 | 1.12910 +j 2.23449e-0... | 25265.4 |
| Mode 3 | 1.15361 +j 6.08843e-0... | 9473.79 |
| Mode 4 | 1.17653 +j 3.37491e-0... | 1.74305e+007 |
| Mode 5 | 1.19272 +j 6.45986e-0... | 9231.77 |
| Mode 6 | 1.22593 +j 6.57058e-0... | 932894. |
| Mode 7 | 1.22794 +j 6.57624e-0... | 9336.19 |
| Mode 8 | 1.25571 +j 7.79531e-0... | 80542.4 |
| Mode 9 | 1.27836 +j 2.89095e-0... | 22109.6 |
| Mode 10 | 1.32295 +j 2.97826e-0... | 22210.2 |
| Mode 11 | 1.32790 +j 4.79650e-0... | 1.38424e+006 |
| Mode 12 | 1.36256 +j 6.17098e-0... | 1.10401e+007 |
| Mode 13 | 1.36375 +j 2.45340e-0... | 27793.0 |
| Mode 14 | 1.37755 +j 2.14777e-0... | 3.20695e+007 |
| Mode 15 | 1.39339 +j 1.69146e-0... | 411892. |
| Mode 16 | 1.41048 +j 1.54322e-0... | 45699.4 |
| Mode 17 | 1.43038 +j 1.49509e-0... | 478361. |
| Mode 18 | 1.44410 +j 8.91517e-0... | 8.09913e+007 |
| Mode 19 | 1.45109 +j 1.78868e-0... | 40563.3 |
| Mode 20 | 1.47155 +j 1.15870e-0... | 6.35002e+006 |

Table .1. Resonating frequency in an Eigen-mode study in the whole coupler structure.

## 5. Conclusion

A novel concept of a variable coupler is introduced and designed. Theoretical analysis has been simulated by HFSS. This variable coupler has the potential capability to deliver up 7 KW continuous RF power. Further optimization can make this coupler compact and suitable for a given configuration in cyro-module. The prototypes of the polarizer will be produced and tested for experimental soon.

## 6. Acknowledgement

We would like to thank Dr ZhengHai Li from SLAC national accelerating laboratory and Dr. Hiroshi Matsumoto of KEK for very useful discussions. This work was supported by Department of Energy Contract No. DE-AC02-76SF00515.

## 7. Reference

1. H. Padamsee, J. Knobloch, and T. Hays, RF Superconductivity for Accelerators. 2nd Edition (Wiley and Sons, New York,NY, 2008).
2. W.-D. Moeller for the TESLA Collaboration, "High Power Coupler For The TESLA Test Facility", Proceedings of the 9th Workshop on the RF Superconductivity, 1999, Santa Fe, V.2, pp.577-581.
3. J. R. Delayen, L.R. Doolittle, T. Hiatt, J. Hogan, J. Mammosser. "An R.F. Input Coupler System For The CEBAF Energy Upgrade Cryomodule." Proceedings of the 1999 Particle Accelerator Conference, New York, 1999. pp1462-1464.
4. S. Belomestnykh, et al., "High Average Power Fundamental Input Couplers for the Cornell University ERL: Requirements, Design Challenges and First Ideas," Cornell LEPP Report ERL 02-8 (September 9, 2002).
5. RF windows

6. E. Snitzer. "Cylindrical dielectric waveguide modes" Journal of the Optical Society of America, Vol. 51, Issue 5, pp. 491-498 (1961)
7. Zaki, K.A. ; Atia, A.E. "Modes in Dielectric-Loaded Waveguides and Resonators. " Microwave Theory and Techniques, IEEE Transactions on V.31 , I.12 ,1982. pp1039 – 1045.
8. Clarricoats, P.J.B ."Properties of dielectric-rod junctions in circular waveguide", Electrical Engineers, Proceedings of the Institution of (Volume:111 , Issue: 1 ). 1964. Pp.43 – 50
9. Clarricoats, P.J.B. ; Taylor, B.C. "Evanescent and propagating modes of dielectric-loaded circular waveguide". Electrical Engineers, Proceedings of the Institution of (Volume:111 , Issue: 12 ) 1964 pp.1951-1956.
10. Walter M. Elsasser. "Attenuation in a Dielectric Circular Rod". J. Appl. Phys. 20, 1193 (1949)
11. Rothwell, E.J. and Frasch, L.L. "Propagation characteristics of dielectric-rod-loaded waveguides".Microwave Theory and Techniques, IEEE Transactions on (Volume:36 , Issue: 3 ) 1988. pp 594 – 600.
12. Optics book
13. S.F.Mahmoud. "Electromagnetics waveguides theory and applications". The Institution of Engineering and Technology (December 1991)
14. A.A. Mostafaa, C.M. Krowneb, K.A. Zakic & S. Tantawid "Hybrid-Mode Fields in Isotropic and Anisotropic Planar Microstrip Structures." Journal of Electromagnetic Waves and Applications Volume 5, Issue 6, 1991